\documentclass{article}

\usepackage{algorithm}
\usepackage[noend]{algpseudocode}
\usepackage{main}

\usepackage[utf8]{inputenc} 
\usepackage[T1]{fontenc}    
\usepackage{hyperref}       
\usepackage{url}            
\usepackage{booktabs}       
\usepackage{amsfonts}       
\usepackage{nicefrac}       
\usepackage{microtype}      
\usepackage{lipsum}
\usepackage{graphicx}
\graphicspath{ {./images/} }

\title{Information Retrieval in long documents: Word clustering approach for improving Semantics}

\author{
  Paul MBATHE MEKONTCHOU \\
  Department of Computer Science\\
  National Advanced School of Engineering - Yaoundé\\
  \texttt{mbathepaul@gmail.com} \\
   \And
  Armel FOTSOH \\
  Sweez - R\&D Department  \\
  Paris - France\\
  \texttt{armel@sweez.io} \\
  \And
  Bernabé BATCHAKUI \\
  Department of Computer Science\\
  National Advanced School of Engineering - Yaoundé\\
  \texttt{bbatchakui@gmail.com } \\
  \AND
  Eddy ELLA \\
  Izysoft - IT Department \\
  Yaoundé - Sydney \\
  \texttt{eddy.insearch@gmail.com} \\
}

\begin{document}
\maketitle
\begin{abstract}
  In this paper, we propose an alternative to deep neural networks for semantic information retrieval for the case of long documents. 
  This new approach exploiting clustering techniques to take into account the meaning of words in Information Retrieval systems targeting long as well as short documents. This approach uses a specially designed clustering algorithm to group words with similar meanings into clusters.  The dual representation (lexical and semantic) of documents and queries is based on the vector space model proposed by Gerard Salton in the vector space constituted by the formed clusters. The originalities of our proposal are at several levels: first, we propose an efficient algorithm for the construction of clusters of semantically close words using word embeddings as input, then we define a formula for weighting these clusters, and then we propose a function allowing to combine efficiently the meanings of words with a lexical model widely used in Information Retrieval. The evaluation of our proposal in three contexts with two different datasets (\textit{SQuAD} and \textit{TREC-CAR}) has shown that is significantly improves the classical approaches only based on the keywords without degrading the lexical aspect.
\end{abstract}




\maketitle

\section{Introduction}
The aim of an Information Retrieval System (IRS) is to identify, in the corpus of documents, those which best meet the information needs contained in the query. Furthermore, organisation's information is contained in documents, which are heterogeneous, unstructured and growing. The content of these documents is usually textual, written in natural language, in unstructured format, which makes automatic processing difficult. These constraints have led to the development of numerous IR models.\\

Traditional information retrieval approaches are most often keyword-based \cite{deo2018survey}. 
For a document to be relevant to a query, it must contain terms identical to those in the query. However, when they carry out searches, users do not always use the same words or the same wording as the content that best answers their queries. Thus, the most relevant documents are not always those containing the same terms as the query (synonymy issue) (\cite{chenaina2022query}, \cite{yadav2021comprehensive}). \\

In view of these difficulties, and in order to improve the performance of traditional search engines, several works targets specific search contexts where domain knowledge can be structured in semantic resources. This is not necessarily suitable when tackling the problem of synonymy, for example, as it would require listing all the words of each language and their synonyms in a kind of dictionary that would be gigantic \cite{mahalakshmi2021art}.  
Moreover, the problem of taking into account the words of the query, from a lexical point of view, remains.\\ 

High performance neural network based IR models have been trained to take into account both the lexicon and the semantics of words in documents. However, they are not at all suitable for long documents.
\\

Our contribution is based on language models allowing to represent words, regardless of language, as a fixed dimension vector: this is the notion of multilingual word embeddings. 
We propose to exploit these representations to group words according to their meaning in clusters. Then, to see each document or query as a set of clusters. Finally, to apply the same matching techniques on clusters rather than on keywords to obtain the similarity score between a query and a document. This allows to improve classical IRSs by taking into account the meaning of words. 
Moreover, in the IRS implementing our proposal, we combine the similarity score obtained previously with the score obtained using \textit{bm25}~\cite{bm25s} to keep at least the same performance when we are in a very lexical IR context. It is important to note that our proposal is independent of the length of the documents in the corpus.
 We evaluate this new IRS with different IR campaigns in order to measure how much it improves the classical IRS. \\\\
 The rest of the paper will be organised as follows: we will first review the main existing approaches to address semantic IR problem. Then, we will present in detail our proposal by highlighting our contribution. After that, we will detail our experiments in different contexts and on different corpora before analysing the results. Finally, we will conclude the paper and present some perspectives opened by this work.

\section{State of the art}
\begin{sloppypar}
   The primary goal of an Information Retrieval System (IRS) is to identify from a corpus of documents those that are relevant to an information need expressed through a query.
   Traditionally, documents and queries are represented through a set of vocabulary keywords \cite{boukhari2020dl}, Then for the matching step, several models have been proposed: the Boolean model (\cite{baezayates}), the Vector model (\cite{salton}, \cite{salton2}) and the Probabilistic model (\cite{rijsbergen}, \cite{robertson1}).
   For lexical IR models, \textit{bm25} \cite{bm25s} is the current state of the art. However, it does not take into account the sense of words.
\end{sloppypar}

\subsection{Semantic Information Retrieval}
\begin{sloppypar}
   Many works have addressed the problem of word meaning in IR. We will group them into three main categories: deep learning approaches based on neural networks, approaches based on semantic resources and approaches based on word embeddings. 
\end{sloppypar}

\subsubsection*{Deep learning approaches}
  \begin{sloppypar}
     Concerning deep learning approaches, the principle is to train large volumes of queries on a document corpus so that the system captures the meaning of words according to their context for future queries (\cite{ref_book1}, \cite{mista_neural}). Some pretrained models of this category such as SBERT\cite{reviews5} have been proposed for performing good zero-shot retrieval for different languages. Since these pretrained models take as input short text sequence, all the context is lost while processing long documents (more than the max token length of the model) because of truncation. Moreover, these approaches generally require large collections of training data and a lot of computational resources to achieve good performance \cite{mista_neural}. 
  \end{sloppypar}

\subsubsection*{Approaches based on semantic resources}
These approaches use semantic resources such as thesauri (\cite{ref_book3}) or ontologies (\cite{ref_book4}, \cite{ref_book5}). The principle is to exploit these resources to annotate documents and queries and then make matches. Most of the existing work has focused on the purely conceptual aspect, so they are effective for corpora with a strong semantic focus or related to a specific domain. However, in a generalist IR context, it is necessary to return precise results by going beyond concepts. Indeed, the majority of corpora are not based on a conceptual resource therefore it is difficult to always use these approaches. 

\subsubsection*{Approaches based on word embeddings}
In these family of approaches (\cite{embb_ir}, \cite{van2017combining}), word meaning is represent by a multi-dimensional vector which is called: \textit{word embedding}. 
\\

In the first sub-family, the document and the query are represented by the average of their weighted word vectors. Weight of each word could be its frequency in the document (TF) or it relative frequency regarding the whole corpus (TF-IDF). 
\\\\
Formally, given a document $d$ with a set of words $W$. It vector is given by 
\begin{displaymath}
\vec{d}= \frac{1}{\sum_{w\in W}^{}weight_{w}} \sum_{w\in W}^{}weight_{w}*\vec{w}
\end{displaymath}

The similarity between a document and the query is computed using the cosine of the angle formed by their respective vectors.
The main issue with this approach is that average of number is not very meaningful.
\\

In the second sub-family, several metrics based on word embeddings are defined to compute query-document relevance score.
Dwaipayan Roy et al. (\cite{reviews1}) use a standard similarity between sets to compute similarity between sets of word embeddings of the query and those of the document.
\\
In the third sub-family, word embedding are used to semantically expand the query in other to take into account synonyms that do not appear in the the query (\cite{reviews3}, \cite{reviews4}).
The weakness of this technique is that it could introduce a lot of ambiguity, since query is expand without using the hold context.

\vspace{0.9cm}

In our proposal, we plan to exploit clustering techniques for grouping words according to their meaning. To the best of our knowledge, no a such proposal has been made. However, clustering has already been used in IR in other contexts.

\subsection{Clustering in Information Retrieval}

 Different hypotheses have been proposed for using clustering in information retrieval (\cite{hypothes_clustering}). The main one is the fact that: documents in the same cluster behave in the same way with regard to relevance to information needs. The assumption is that if there is a document in a cluster that is relevant to a search query, then it is likely that other documents in the same cluster are also relevant. 

\subsubsection*{Grouping of search results}
In this use case, the aim is to display the search results not as a list of candidates, but so that returned documents that are similar (i.e. are in the same cluster) appear together. This is particularly useful for polysemous query terms. The works of \cite{finalias} are an illustration of this use case.

\subsubsection*{Clustering in Image Retrieval}
Clustering has also been used in image processing, i.e. image segmentation with the aim of retrieving important information from images. \cite{article999},\cite{article888} propose two main clustering methods for grouping pixels, a hierarchical clustering method and a partitional method. Since partitional clustering is computationally better, they put special emphasis on it in the perspective of methods belonging to this class.

\begin{figure*}[h]
    \centering
    \includegraphics[scale=.5]{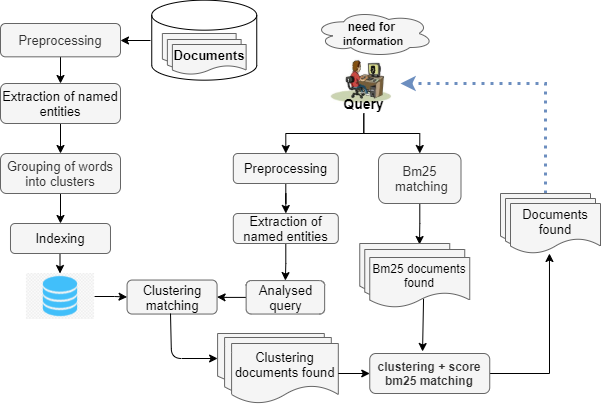}
    \caption{Architecture of the clustering approach}
    \label{fig:architecture_clustering}
\end{figure*}

\section{Proposals}
\begin{sloppypar}
     We propose to use clustering to group words according to their meaning. For this purpose, we use word embeddings as a semantic representation of words and as an entry point for clustering. The advantage of this approach is that these word-embeddings are already pretrained and multilingual, which makes it possible to solve the word meaning problem without computational constraints. We also combine this approach with the \textit{bm25} probabilistic model in an original processing chain. The aim of this combination is to guarantee a high degree of robustness even for searches where the meaning is not very important.
\end{sloppypar}

\subsection{Architecture of our proposal}
\label{architecture}
The processing chain describing our IRS is illustrated in Figure \ref{fig:architecture_clustering}. It consists of a set of modules to process the documents and the query before matching them.
\begin{itemize}
    \item For each document and query, we annotate it in order to identify Named Entities that do not necessarily carry meaning. This makes it possible to distinguish key words from those for which a purely lexical matching is enough.
    
    \item Then, the words of the documents are grouped into clusters according to whether their meanings are close or not. For this purpose, we use the vector representations of multilingual word embeddings. Named entities and rare words are alone in their clusters to reduce ambiguity and noise.
 
    \item Each document is thus represented as a vector in the cluster space. The weighting function of each of them is one of our contributions. Similarly, the query is also represented in the same vector space.
    \item From the representation of documents and queries, we compute a similarity score using the cosine function. Similarly, we compute a purely lexical score with \textit{bm25} which we combine with the previous one to determine the relevant documents. 
\end{itemize}

\subsection{Text pre-processing} \label{processing}
The text of each document and query is processed through a set of operations to remove anything that might be noise: hashtags, HTML tags, stop-words, etc.

\subsection{Recognition of named entities}
Named entity recognition allows us to label words in the text in order to improve the algorithm for grouping words into clusters. Indeed, named entities do not necessarily have synonyms and can be a source of ambiguity. In fact, with regard to the way word embeddings are trained, proper names can be represented by close vectors because they are generally used in the same context ("Axtérix" and "Obelix"). The same applies to place names and other names. Recognising named entities here avoids these ambiguities, which would have had a negative impact on our IRS. Thus, named entities are alone in their cluster and do not overlap other words.
 
\subsection{Word grouping in clusters}
\label{construction_clusters_}
The annotated text is tokenized to build the vocabulary of the corpus. The resulting list of tokens is used to build clusters. 
Moreover, for a generic IRS, some constraints must be taken into account:
\begin{itemize}
    \item the number of clusters should not be known in advance. Indeed, as for classical IRS, the size of the vocabulary is not fixed;
    \item the algorithm should be incremental in order to allow new words to be added to the clusters without having to globally modify the clusters already formed;
    \item The centroid must be fixed to prevent a cluster from taking into account words with different meanings;
    \item the clusters formed must be linearly independent from a semantic point of view for them to form a vector space; 
    \item Rare words (RW) and Named Entities (NE) should be alone in their clusters,
\end{itemize}
In view of these different constraints, we propose an algorithm for the construction of clusters, based on the Single Pass Algorithm \cite{gupta2004genic}.

\label{algori}
\begin{algorithm}
\caption{Grouping words from the collection into clusters}
\begin{algorithmic}[1]
\Require $\epsilon \geq 0$ 
\State $Cluster \;\; new\_cluster$ \Comment{Create a new cluster}
\If{$clusters.length=0\;\;or\;\; w.label="NE-RW"$}
    \State $new\_cluster.centroid \gets w.vecteur$ 
    \State $add\_word(new\_cluster.words,  \; w.text)$  
    \State $add\_cluster(clusters, \; new\_cluster)$ 
   
\Else
    \State $cluster\_index \gets None$
    \State $cluster\_index  \gets find\_closest\_cluster(clusters,w,\epsilon)$
   \If{cluster\_index = None}
        \State $new\_cluster.centroid \gets w.vecteur$
        \State $add\_word(new\_cluster.words.add,w.text)$  
        \State $add\_cluster(clusters,\;new\_cluster)$
    \Else
        \State $add\_word(clusters[cluster\_index].words,\;w.text)$
    \EndIf
\EndIf
\end{algorithmic}
\end{algorithm}

\subsubsection*{Description}
\begin{itemize}
    \item If the list of clusters is empty or processed word is Named Entity or rare word, a new cluster is created.
    \item otherwise, we look for the cluster closest to the word with the $\epsilon$ threshold.
    This by comparing the embedding of this word with each cluster centroid and check if the similarity score is lower than $\epsilon$.
    \item If the word does not belong to any cluster, we create a cluster.
    \item If the word fits into a cluster, it is added to the list of words in that cluster without changing the centroid.
   \item the functions \textit{add\_word(L,w) } and \textit{add\_cluster(L,w) } allow a word $w$ to be added to the list words or cluster respectively.
\end{itemize}

By looking for the closest cluster to the word rather than the first cluster in which the word could be added, ambiguity is reduced by ensuring that a word does not belong to more than one cluster.

\subsubsection*{Selection of the $\epsilon$ threshold}
$\epsilon$ is determined by selecting N pairs of words from the target language, which are known to be synonymous. Using the word-embeddings of these words, the cosine score between them is determined and the parameter $\epsilon$ is the average of the scores obtained over all pairs.

\subsubsection*{Complexity}
The addition of a word requires in the worst case to go through all the clusters to find the closest one. And for each round of the loop, it is necessary to calculate the distance between the centroid of the cluster and the vector associated with the word. The algorithm would thus run in $O(n*\beta)$ where $n$ is the number of clusters at the time of the insertion of the word and $\beta$ the complexity to calculate the cosine between the centroid and the word vector.

\begin{sloppypar}
  In our implementation, we use \textit{fastText} embeddings for building the clusters. 
  These clusters are then save in Elasticsearch with their centroid vectors.
\end{sloppypar}

\subsection{Representation of a document in the cluster space}
According to the constraint on the construction of the clusters, they are linearly independent and form a vector space.
Therefore, we represent each document in this space. \\
Assuming that we have $n$ clusters built from the documents of the corpus treated, the document $d_j$ is given by the vector:
$$\vec{d_j}= (\alpha _1^j,\alpha _2^j,\ \dots \ \alpha_n^j)$$ 
where $\alpha_i^j$ the weight of cluster $i$ in the document $d_j$.

\subsubsection*{Calculation of the weight $\alpha_i^j$ of cluster $i$ in document $j$}
For the calculation of the $\alpha_i^j$ weight, we use the principle of the \textit{TF-IDF} weighting which we adapt in an original way to the case of clusters. The weight $\alpha_i^j$ of the cluster $i$ in the document $j$ is calculated with the following formula:
\begin{displaymath} 
\alpha_i^j = \beta\log(1+F_i^j)\log(\frac{N}{N_i+1})
\end{displaymath}
\begin{itemize}
    \item $\beta$ is a factor that represents the proportion of words in cluster $C_i$ that belong to document $d_j$
    \item $F_i^j$ is the frequency of cluster $C_i$ in document $d_j$, it is equal to the sum of the frequencies of each word in document $d_j$
    \item $N$ is the total number of documents in the corpus
    \item $N_i$ is the number of documents that have at least one word from cluster $C_i$
\end{itemize}
\subsubsection*{Justification}
We added the $\beta$ factor to increase the weight of a cluster if it contains several distinct words.
By calculating the frequency of a cluster as the sum of the frequencies of the words it contains, we thus favour clusters that have several words that appear several times in a document. Indeed, the objective here is to minimize the weight of clusters that have too many words and share few words with a document, as these clusters will create ambiguity. Let us suppose that we have two clusters $C_1$ and $C_2$ and a document $d$. The cluster $C_1$ has 10 words and the cluster $C_2$ has 5. Assuming that the cluster $C_1$ contains 3 words in document $d$ and $C_2$ contains also 3 words in that document, by calculating the factor $\beta$ of each, we obtain $\frac{3}{10}$ and $\frac{3}{5}$ respectively for each of the clusters. Our idea here is to minimize the weight of clusters that have too many words and share few words with a given document, as these clusters will create ambiguity.

\subsection{Representation of the query}
The query is pre-processed exactly as documents. The pre-processed and annotated text is then analyzed to be represented in the same cluster space as the documents.
Thus a query $q$ is represented as follow:

$$\vec{q}= (q_1, q_2,\ \dots \ q_n)$$ 
where $q_i$ the weight of cluster $i$ in the query $q$.
\\
\\
A cluster $C_i$ is highlighted in the query vector $q$ if there is a word of this cluster which is in those of the query or if a word of the query can be inserted into that cluster. In other words, if there is a word $w$ in $q$ so that $\cos(\overrightarrow{C_i.centroid},\overrightarrow{w.vector})<\epsilon$.
Where $\epsilon$ is the similarity threshold for a word to belong to a cluster.
\\
Using this approach, the IR system will be able to provide relevant results even for query  with words not appearing in the corpus of documents. 
\\
Consider that we are interested in computing the weight $q_i^w$ of the cluster $C_i$ regarding a word $w$ of the query.

\begin{displaymath} 
    q_i^w =\left\{ \begin{array}{rr}
    \gamma \ \ \ \ \ \ if\  w \in\  C_i.words\\
    f(g(\overrightarrow{C_i.centroid},\overrightarrow{w.vector}))\ \  else\\  
    \end{array}
    \right. \
\end{displaymath}
\\
with
\begin{displaymath} 
g(\vec{u},\vec{v})=
  \left\{ \begin{array}{ccc}
   0\ \  si\ \  \cos{(\vec{u},\vec{v})}>\epsilon\\
  cos{(\vec{u},\vec{v})}\ \  si\ \  \cos{(\vec{u},\vec{v})}\leq\epsilon\\
  \end{array}
 \right. \
\end{displaymath}
 and
\begin{displaymath}
f(x)=\frac{\gamma}{\epsilon}(\epsilon-x)
\end{displaymath}

$\gamma \in \mathbb{R}$ such as $\gamma>0$ represents the maximum weight given to a cluster in query representation.
$\overrightarrow{C_i. centroid}$ is the centroid vector of the cluster $C_i$, and $\overrightarrow{w.vector}$ the vector associated to the word $w$. 

\begin{sloppypar}
The function $g$ allows to calculate the semantic proximity of a word with a cluster regarding the centroid.
$f$ reduce the weight for word that do not belong to a cluster but can be semantically close.
\end{sloppypar}

Finally, the weight $q_i$ of the cluster $C_i$ in the query vector q is defined by 
\begin{displaymath}
q_i = \sum_{w\in q.words}^{}q_i^w
\end{displaymath}
$q.words$ is the set of words in query q.

\subsection{Document-query matching}
The matching between a given document and a query is done by calculating the cosine of the angle between the document vector and the query vector in the space formed by the clusters. For a document $d$ given by :
$$\vec{d}= (\alpha_1,\alpha_2,\ \dots \ \alpha_n)$$ 
and a query $q$ represented by :
$$\vec{q}= (q_1, q_2,\ \dots \ q_n)$$ 
The document \textit{RSV} (Retrieval Status Value) for the query is calculated with the formula :
\begin{displaymath}
RSV(d,q)=cos(\vec{d},\vec{q})=\frac{\sum_{i=1}^{n}\alpha_i.q_i}{\sqrt{\sum_{i=1}^{n}(\alpha_i)^2}\sqrt{\sum_{i=1}^{n}(q_i)^2}}
\end{displaymath}

\subsection{Cluster-keyword combination}
At the end, we combine the \textit{clustering} approach with the best keyword-based approach: \textit{bm25}\cite{bm25s}.
For this purpose, we use a formula derived from the voting system of \textit{Borda}.
Thus, for a given query $q$, the global score of a document $d$ having a rank $n$ and a score $s_{clustering}$ for the clustering approach and a rank $m$ and a score $s_{bm25}$ for \textit{bm25} is defined as follow:
\begin{displaymath}
RSV(d,q)=(N-n)*s_{clustering}+(N-m)*\log(1+s_{bm25})
\end{displaymath}
where $N$ is the number of documents returned and $s_{bm25}$ represents the \textit{bm25} score normalized in the interval $[min,max]$. $min$ and $max$ represent respectively the minimum and maximum score of the \textit{clustering} approach for the query.

\section{Evaluations}
\subsubsection*{Evaluation goals}
\begin{sloppypar}
  We implement our proposal in other to evaluate its performance on two campaigns widely used in the state of the art. 
  Our baselines consists of two models: 
  \begin{itemize}
      \item \textit{bm25} which is presented as the reference in terms of keywords and used in several search engines
      \item \textit{fasttext avg} which is in our implementation based on fasttext pretrained word embeddings \cite{fasttext}.
  \end{itemize}
\end{sloppypar}

The objective of our evaluation is threefold:

\begin{itemize}
    \item We want to know how our combined model can improve \textit{bm25}.
    \item Also, we want to study whether these improvements do not impact the performance on purely lexical corpora.
    \item We want to compare our model with a naive word embedding average approach.
\end{itemize}
For this purpose, we use two data sets: \textit{SQuAD}\cite{online2} and \textit{TREC-CAR}\cite{online4}.
\\
\subsubsection*{IRS implementation \& Metrics}
\begin{sloppypar}
    In our implemented IR system, we use the well-known XLM-RoBERTa~\cite{xlm-roberta} of the \textit{transformers} package and bot lexical and cluster-vector for each document are store in Elasticsearch.
\end{sloppypar}

\begin{sloppypar}
    We will evaluate our results by analysing several metrics widely used for this purpose in the state of the art~\cite{harman2011information}: Precision, Recall, MAP, R-PREC and MRR.
    We will also compare the Precision and Recall of each target model for the top-k result set of queries. We use the paired Student's t-test to be sure that the Precision series (Recall respectively) of the different models are not related.
\end{sloppypar}

\subsection{Evaluation on SQuAD}
We used the train 2.0 version of the dataset which contents of 51324 queries (questions) covering 10698 documents (paragraphs).
We evaluate two scenarios:
\begin{itemize}
    \item In the first, we use the dataset as it is.
    \item In the second, we rephrase the questions by randomly replacing some words with synonyms.
\end{itemize}
\subsubsection{Scenario 1: SQuAD without reformulation}
For this first scenario the context is assumed to be lexical, i.e. the words in the query are probably the same as those used in the document.

\begin{sloppypar}
  The table  \ref{tab:global_squard_sans_reform} presents the global performances on the corpus. Overall performance of our combined model is better than \textit{bm25} and \textit{fasttext avg} for all evaluated metrics. However, clustering approach alone is slightly less efficient that \textit{bm25}. This is due to the fact that the analyzed corpus is lexical and grouping word into clusters may sometimes introduce some noises while word sense is not exploited. These noises are offset with the combination function.
\end{sloppypar}

\begin{figure}[H]
        \centering
        \includegraphics[scale=.55]{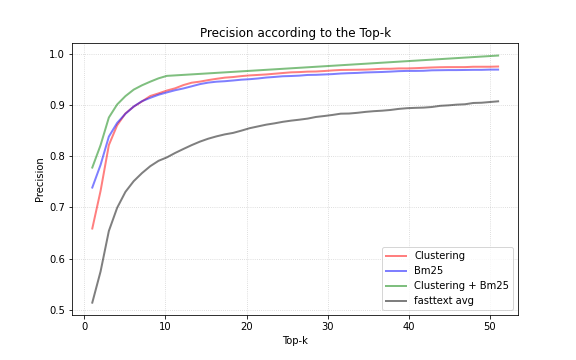}
        \caption{precision according to the reformulated Top-50 SQuAD}
        \label{fig:prec_sans_reform_squard}
        \centering
         \includegraphics[scale=.55]{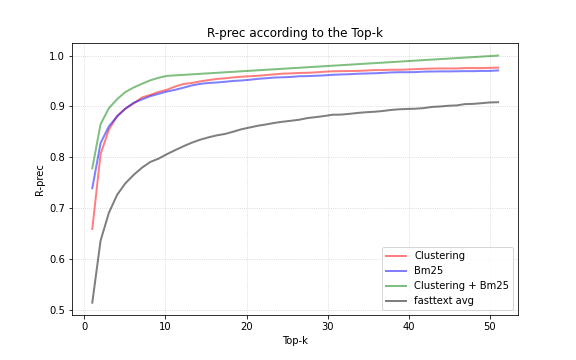}
        \caption{R-Prec as a function of the reformulated Top-50 SQuAD}
        \label{fig:rappel_sans_reform_squard}
\end{figure}

Figures \ref{fig:prec_sans_reform_squard} and \ref{fig:rappel_sans_reform_squard} show the evolution of precision and recall respectively as a function of Top-k ($k \in [1, 50]$).

\begin{sloppypar}
 We find that the precision of the fasttext avg and clustering-based approach is slightly lower than the one of \textit{bm25}, but the combination of clustering and \textit{bm25} has the highest precision.
\end{sloppypar}

\begin{sloppypar}
  The recall here has the same tendency as the precision. Indeed, for words of the query having synonyms in some documents, these will be highlighted with the clustering approach. However the expected documents are those containing almost the same words as in the query, hence the drop in performance. 
\end{sloppypar}

\begin{table}[H]
  \begin{center}
    \caption{Overall performance on SQuAD}
    \begin{tabular}{|c|c|c|c|}
      \hline
      \textbf{} & 
      \textbf{MAP} &
      \textbf{R-Prec} &
      \textbf{MRR}  \\
      \hline
      \hline
      \textit{bm25} & 0.867 & 0.738 & 0.806 \\ \hline
      fasttext avg & 0.706 & 0.513 & 0.614 \\ \hline
       Clustering & 0.853 &0.658 & 0.763  \\ \hline
      Clustering+\textit{bm25} & \textbf{0.896} & \textbf{0.777} &  \textbf{0.840}\\ \hline
    \end{tabular}
    \label{tab:global_squard_sans_reform}
  \end{center}
\end{table}

\subsubsection{Scenario 2: SQuAD with reformulation}
\begin{sloppypar}
    In this second scenario, we evaluate the performance of our IRS in a context where the queries are reformulated in the \textit{SQuAD} dataset. Indeed, some words of the query are randomly replaced by synonyms (same meaning in the context of use). 
    These synonyms are selected using Wordnet~\cite{wordnet}.
\end{sloppypar}

\begin{table}[h]
  \begin{center}
    \caption{Overall performance (MAP, R-Prec, MRR)}
    
    \begin{tabular}{|c|c|c|c|}
      \hline
      \textbf{} & 
      \textbf{MAP} &
      \textbf{R-Prec} &
      \textbf{MRR}  \\
      \hline
      \hline
       \textit{bm25} & 0.676 & 0.493 &  0.592 \\ \hline
        fasttext avg & 0.642 & 0.433 &  0.548 \\ \hline
       Clustering & \textit{0.785} & \textit{0.570} & \textit{0.687} \\ \hline
      Clustering+\textit{bm25} & \textbf{0.831} & \textbf{0.660} & \textbf{0.754} \\ \hline
    \end{tabular}
    \label{tab:perfom_global_squard_reform}
  \end{center}
\end{table}

\begin{sloppypar}
  The table \ref{tab:perfom_global_squard_reform} presents overall results for the evaluation of the different approaches.
  Also in this scenario, our combined approach achieves the best results for all metrics. However, unlike the previous scenario, the clustering only approach performs better than \textit{bm25}. This can be explained by the fact that it makes it possible to better take into account word meaning, which \textit{bm25} does not do at all. Therefore, the use of a synonyms rather that a work from the corpus should not be an issue for retrieving the correct document.
\end{sloppypar}

\begin{figure}[H]
        \centering
        \includegraphics[scale=.55]{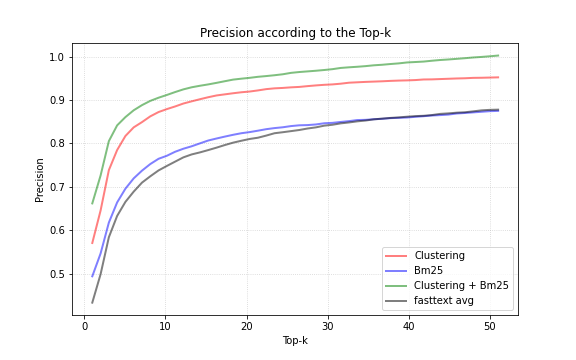}
        \caption{Precision according to the reformulated Top-50 SQuAD}
        \label{fig:prec_reform_squard}
   
    \vfill%
    
        \centering
         \includegraphics[scale=.55]{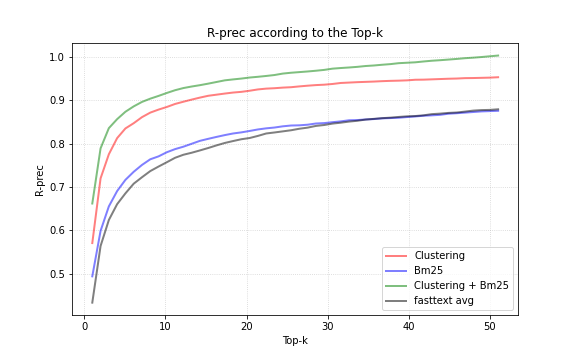}
        \caption{R-Prec as a function of the reformulated Top-50 SQuAD}
        \label{fig:rappel_reform_squard}
\end{figure}

\begin{sloppypar}
  Figures \ref{fig:prec_reform_squard} and \ref{fig:rappel_reform_squard} show the evolution of precision and recall as a function of the Top-50.
\end{sloppypar}

\begin{sloppypar}
   The analysis of these curves show that even if we are on a semantic corpus, the \textit{fasttext avg} approach is the worst. This illustrates that the weighed average of embeddings does not necessary make it possible to aggregate the sense of the words of a document. Moreover, the two approaches based on clustering clearly stand out from the others. 
\end{sloppypar}

\subsection{Evaluation on TREC-CAR}
\begin{sloppypar}
   We also evaluated our model and fasttest avg approach on the test version of a part of the corpus \textit{TREC-CAR} which is constituted of 76525 documents and 19525 queries randomly selected in the global test set.
\end{sloppypar}

\begin{table}[H]
  \begin{center}
    \caption{Global performance (MAP, R-prec, MRR)}
    
    \begin{tabular}{|c|c|c|c|}
      \hline
      \textbf{} & 
      \textbf{MAP} &
      \textbf{R-Prec} &
      \textbf{MRR}  \\
      \hline
      \hline
      \textit{bm25} & 0.575 & 0.330 &  0.687 \\ \hline
      fasttext avg & 0.423 & 0.187 &  0.482 \\ \hline
      Clustering & 0.471 & 0.237 & 0.555  \\ \hline
      Clustering+\textit{bm25} & \textbf{0.608} & \textbf{0.355} &  \textbf{0.723} \\ \hline
    \end{tabular}
    \label{tab:global_trac_car}
  \end{center}
\end{table}

\begin{sloppypar}
  The table \ref{tab:global_trac_car} presents the overall performance on the corpus.
  As for the previous experiments on \textit{SQuAD}, we find that the combination of \textit{clustering} and \textit{bm25} is the best. 
  Indeed, The TREC-CAR dataset is highly lexical since queries are Wkipedia title and documents section content. 
  This confirms the fact that the \textit{clustering} approach makes it possible to boost \textit{bm25} from a semantic point of view without impacting the initial performances.
\end{sloppypar}

\begin{figure}[h]
    
        \centering
        \includegraphics[scale=.55]{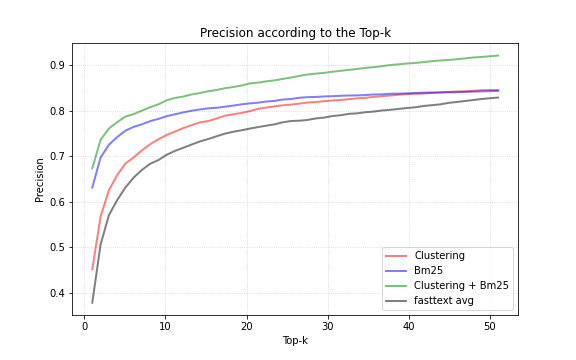}
        \caption{Precision according to TREC-CAR Top-50}
        \label{fig:prec_trac_car}
    
    \vfill%
   
        \centering
         \includegraphics[scale=.55]{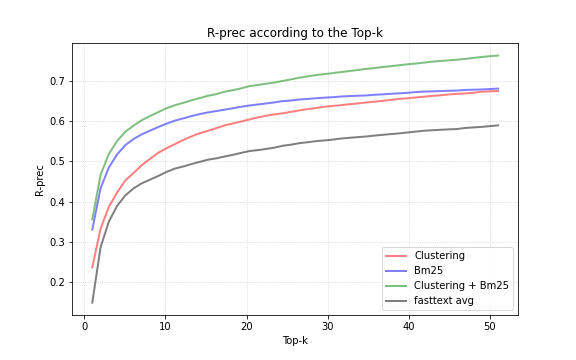}
        \caption{R-Prec according to TREC-CAR Top-50}
        \label{fig:rapel_trac_car}
    
\end{figure}

\begin{sloppypar}
  The figures \ref{fig:prec_trac_car} \ref{fig:rapel_trac_car} show precision and recall as a function of the Top-50, respectively.
\end{sloppypar}

\begin{sloppypar}
    As we observe previously, \textit{fasttext avg} performs worst on the dataset. The \textit{clustering} only is strongly close to the \textit{bm25} from the Top-20. 
\end{sloppypar}

\subsection*{Conclusion on the evaluations}
\begin{itemize}
    \item The combination of \textit{bm25} and \textit{clustering} effectively improves \textit{bm25} by taking into account word sense. This remains true even if we are on a highly lexical or the highly semantic Information Retrieval context.
    \item \textit{clustering} approach alone is not very effective on a highly lexical corpus.
    \item \textit{clustering} approach performs better than word embeddings average one regardless of the retrieval context. 
\end{itemize}

\section*{Conclusion}
Our objective was to propose a solution to efficiently combine word sense and keywords in order to improve the performance of Information Retrieval Systems. We propose a generic processing chain exploiting the grouping of words into semantic clusters. This means that each cluster must include words that have similar meanings. We exploit multilingual word-embeddings for this purpose. The clusters thus constituted a vector space and each document or query is represented in it. We propose original functions to compute the weights of each cluster in these vectors. Finally, we evaluate the semantic score between a document and a query by determining the cosine of the angle between their vectors. This score is combined with the one determined with the \textit{bm25} approach to take into account simultaneously the meaning of the words and the keywords.
The experimentation of this processing chain on two datasets shows the efficiency of our proposal in taking into account the meaning and the keywords. It is important to note that this proposal remains just as effective in a purely lexical search context. This result is interesting in the sense that it really illustrates the flexibility of our system.
However, a perspective could be to use our IRS in front of a Question-Answering system in order to filter the paragraphs likely to contain the answer before passing them to a neural network inference model. This would greatly optimise the processing time when we are sure that the short list of paragraphs to be processed already takes into account the meaning of the questions asked.

\bibliographystyle{ACM-Reference-Format}
\bibliography{main}

\end{document}